\documentclass{article}
\usepackage[utf8]{inputenc}
\usepackage[final]{gpsmdms_2022}
\usepackage{lipsum}
\usepackage{graphicx}
\usepackage{amsfonts}
\usepackage[numbers]{natbib}
\usepackage[hyphens]{url}
\usepackage[hidelinks]{hyperref}
\hypersetup{breaklinks=true}
\usepackage{amsmath}
\usepackage{caption}
\usepackage{subcaption}
\usepackage{algorithm}
\usepackage{algorithmic}

\title{Deep Gaussian Process-based Multi-fidelity Bayesian Optimization for Simulated Chemical Reactors}
\author{Tom Savage\\
Department of Chemical Engineering\\
Imperial College London\\
London,UK\\
\texttt{trs20@ic.ac.uk} 
\And 
Nausheen Basha\\
Department of Chemical Engineering\\
Imperial College London\\
London,UK\\
\texttt{nausheen.basha@imperial.ac.uk}\\
\And
Antonio Del Rio Chanona\\
Department of Chemical Engineering\\
Imperial College London\\
London,UK\\
\texttt{a.del-rio-chanona@imperial.ac.uk}\\
\And
Omar Matar\\
Department of Chemical Engineering\\
Imperial College London\\
London,UK\\
\texttt{o.matar@imperial.ac.uk}
}

\date{August 2022}

\begin{document}

\maketitle

\begin{abstract}
New manufacturing techniques such as 3D printing have recently enabled the creation of previously infeasible chemical reactor designs.
Optimizing the geometry of the next generation of chemical reactors is important to understand the underlying physics and to ensure reactor feasibility in the real world. 
This optimization problem is computationally expensive, nonlinear, and derivative-free making it challenging to solve.
In this work, we apply deep Gaussian processes (DGPs) to model multi-fidelity coiled-tube reactor simulations in a Bayesian optimization setting. 
By applying a multi-fidelity Bayesian optimization method, the search space of reactor geometries is explored through an amalgam of different fidelity simulations which are chosen based on prediction uncertainty and simulation cost, maximizing the use of computational budget.
The use of DGPs provides an end-to-end model for five discrete mesh fidelities, enabling less computational effort to gain good solutions during optimization.
The accuracy of simulations for these five fidelities is determined against experimental data obtained from a 3D printed reactor configuration, providing insights into appropriate hyper-parameters.
We hope this work provides interesting insight into the practical use of DGP-based multi-fidelity Bayesian optimization for engineering discovery.

\end{abstract}

\section{Introduction}
The fluid flow within a reactor greatly influences product quality and is highly dependent on the reactor geometry. 
Plug-flow conditions, corresponding to a product distributions with low variance, are desired.
We study coiled tube reactors, which have been shown to demonstrate promising plug flow performance in previous computational and experimental studies \citep{Mansour2020a,Mansour2017,McDonough2019a,McDonough2019b,Lira2022}.

Computational fluid dynamics (CFD) simulations of coiled-tube reactors are expensive due to complex flow characteristics, and gradient information is practically unavailable. The resulting expensive black-box optimization problem is analogous to hyper-parameter optimization \citep{HPO,larson_menickelly_wild_2019,Feurer2019,NIPS2011_86e8f7ab}, with chemical discovery \citep{Jorayev2022,Mateos2019}, and engineering design \citep{Lam2018,Morita2022,Daniels2019} also examples of domains where expensive black-box optimization problems are formulated.
The problem can be stated as
\begin{align}
    x^* = \mathop{\mathrm{argmax}}_{x\in \mathcal{X}\subset \mathbb{R}^d} f(x). \label{BB}
\end{align}

In certain situations, computational expense can be traded off with accuracy via one or multiple fidelity parameters. 
Examples include training epochs \citep{PMFBO,Schmucker2020} in the context of hyper-parameter optimization, mesh fidelities in the context of finite element analysis \citep{Huang2006}, or combining real-time measurements and predictions in industrial processes \citep{panos}.
Including fidelity control within a Bayesian optimization framework enables optimization with fewer computational resources whilst gaining a `high fidelity' solution \citep{BOAH,march_willcox_wang_2011}. Equation \ref{BB} then becomes 
\begin{align}
    x^* = \mathop{\mathrm{argmax}}_{x\in \mathcal{X}\subset \mathbb{R}^d} f(x,s)
\end{align}

where potentially M different fidelities, $s\in\mathbb{R}^M$ become controllable parameters.

\begin{figure}[t!]
    \centering
    \includegraphics[height=4cm]{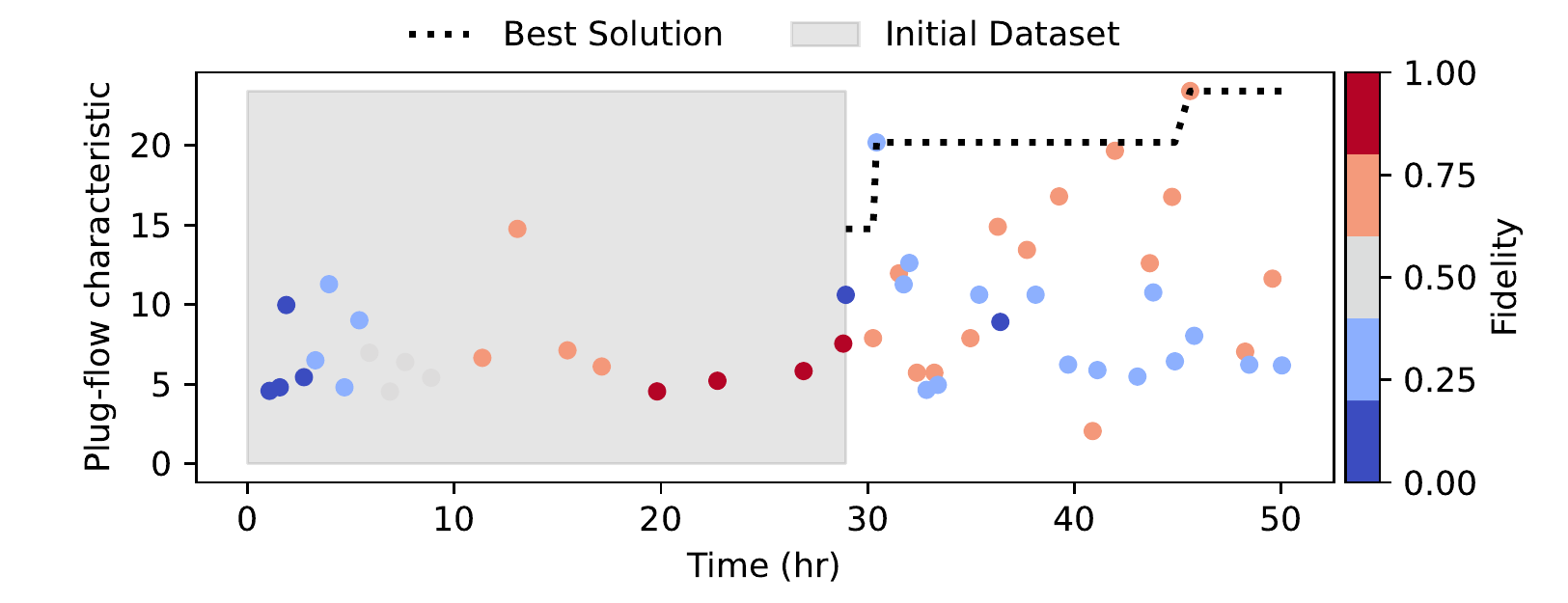}
    \includegraphics[height=4cm]{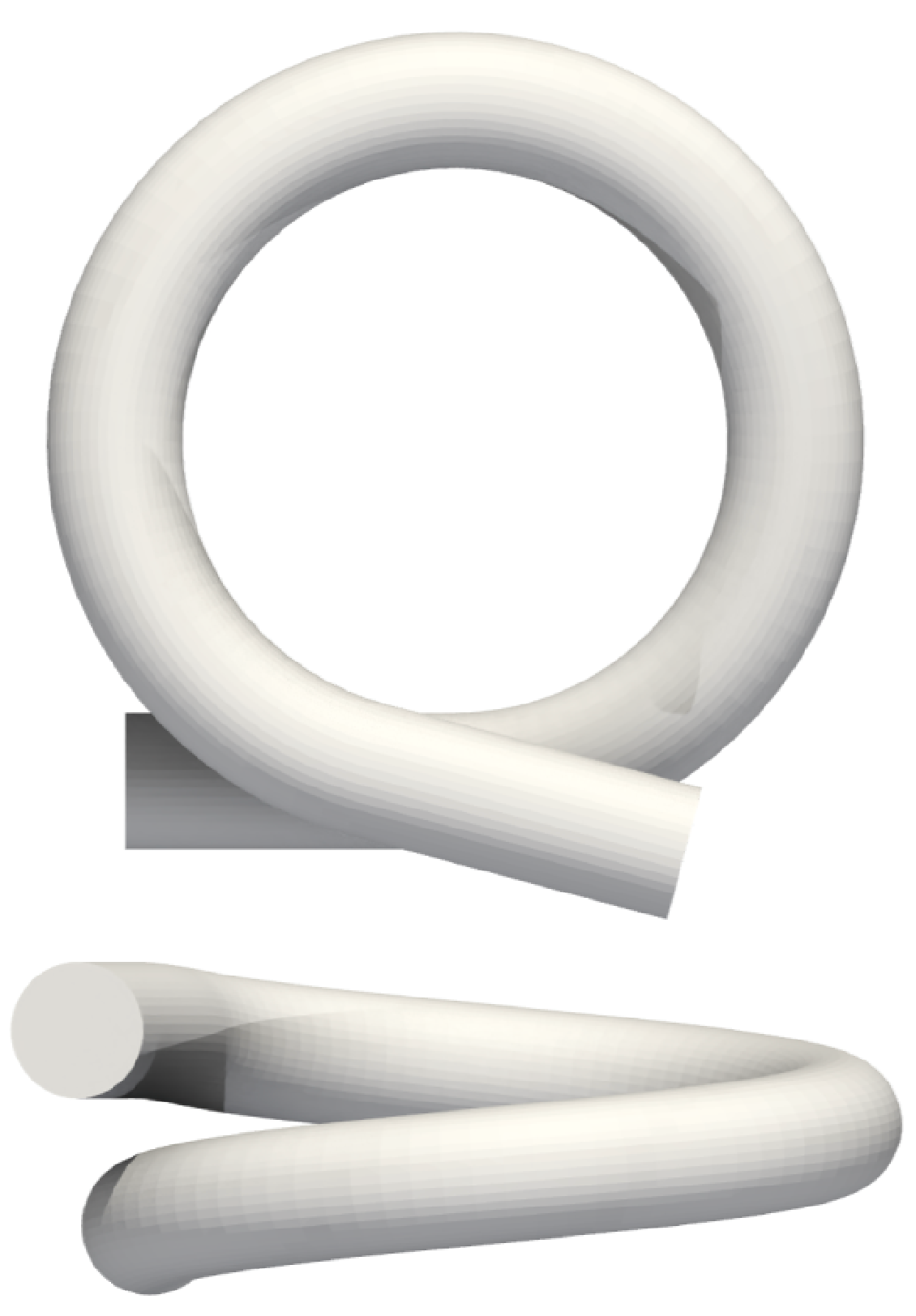}
     \caption{Left: The progression of multi-fidelity Bayesian optimization using DGPs as a multi-fidelity model. Multiple different fidelity levels are selected for evaluation throughout optimization, reducing optimization time. Right: the best coil geometry which has a relatively large coil radius and low pitch.}
     \label{main_results}
\end{figure}

\textbf{Contribution:} In this work, we present the novel real-world problem of optimizing the geometry of a coiled tube reactor to maximize the plug-flow performance.
We apply a state-of-the-art deep GP-based multi-fidelity Bayesian optimization algorithm to identify novel reactor configurations using an amalgam of different fidelity simulations, modeled using a DGP. Figure \ref{main_results} demonstrates how our approach takes advantage of lower fidelity simulations.
Our approach contains no additional hyper-parameters when compared to standard UCB Bayesian optimization.
Having identified an optimal geometry  we investigate the physical insights to inform future design of pulsed-flow coiled tube reactors. 

\section{Method}
By applying \textsc{DGPs} within a Bayesian optimization framework, we enable an end-to-end model of all fidelities.
A more accurate model of higher-fidelities should enable the optimization procedure to make more evaluations at lower, less expensive fidelities.
\subsection{Model fidelities}
\begin{figure}[htb!]
    \centering
    \begin{subfigure}[b]{0.45\textwidth}
         \centering
         \includegraphics[width=0.32\textwidth]{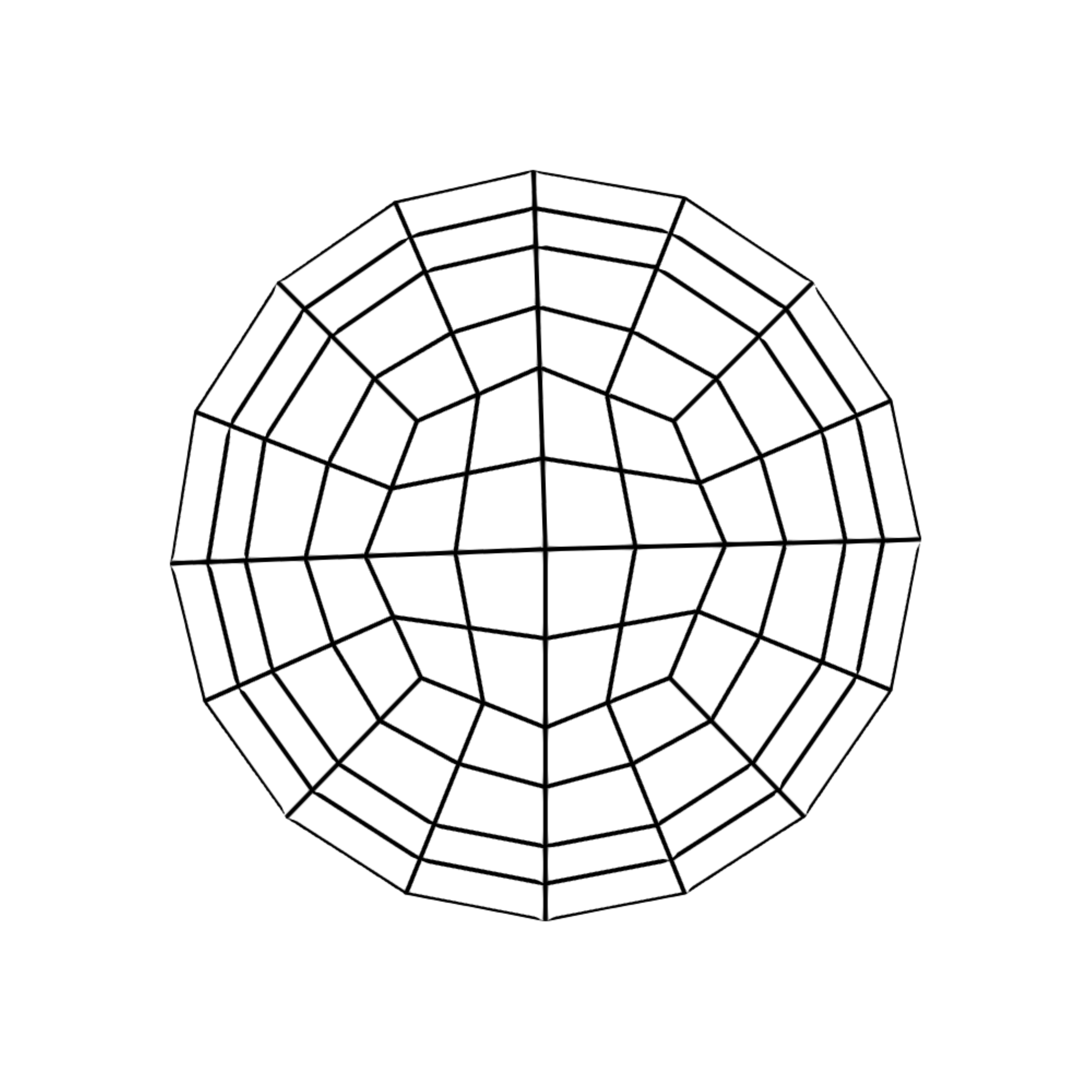}
         \includegraphics[width=0.32\textwidth]{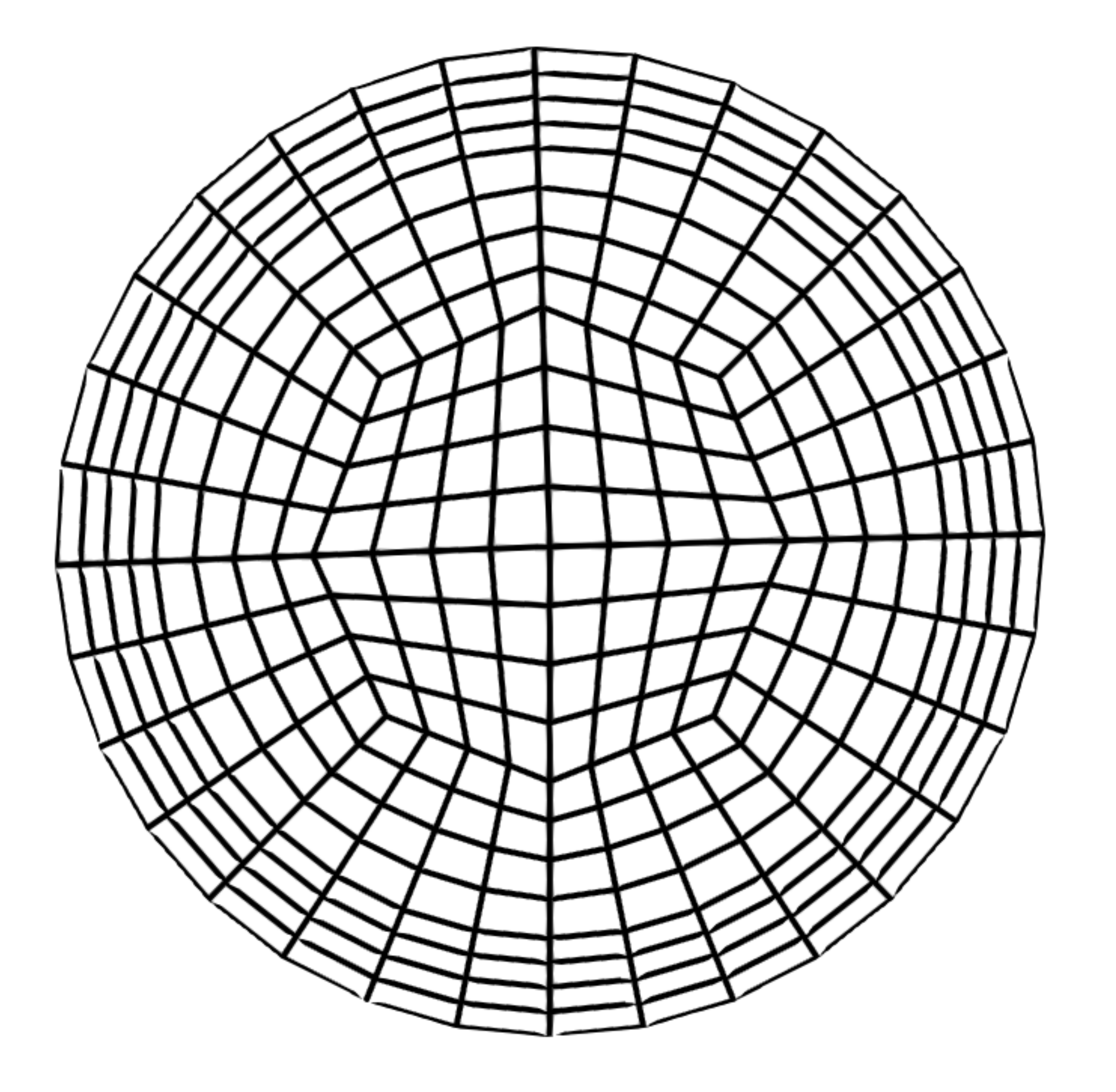}
         \includegraphics[width=0.32\textwidth]{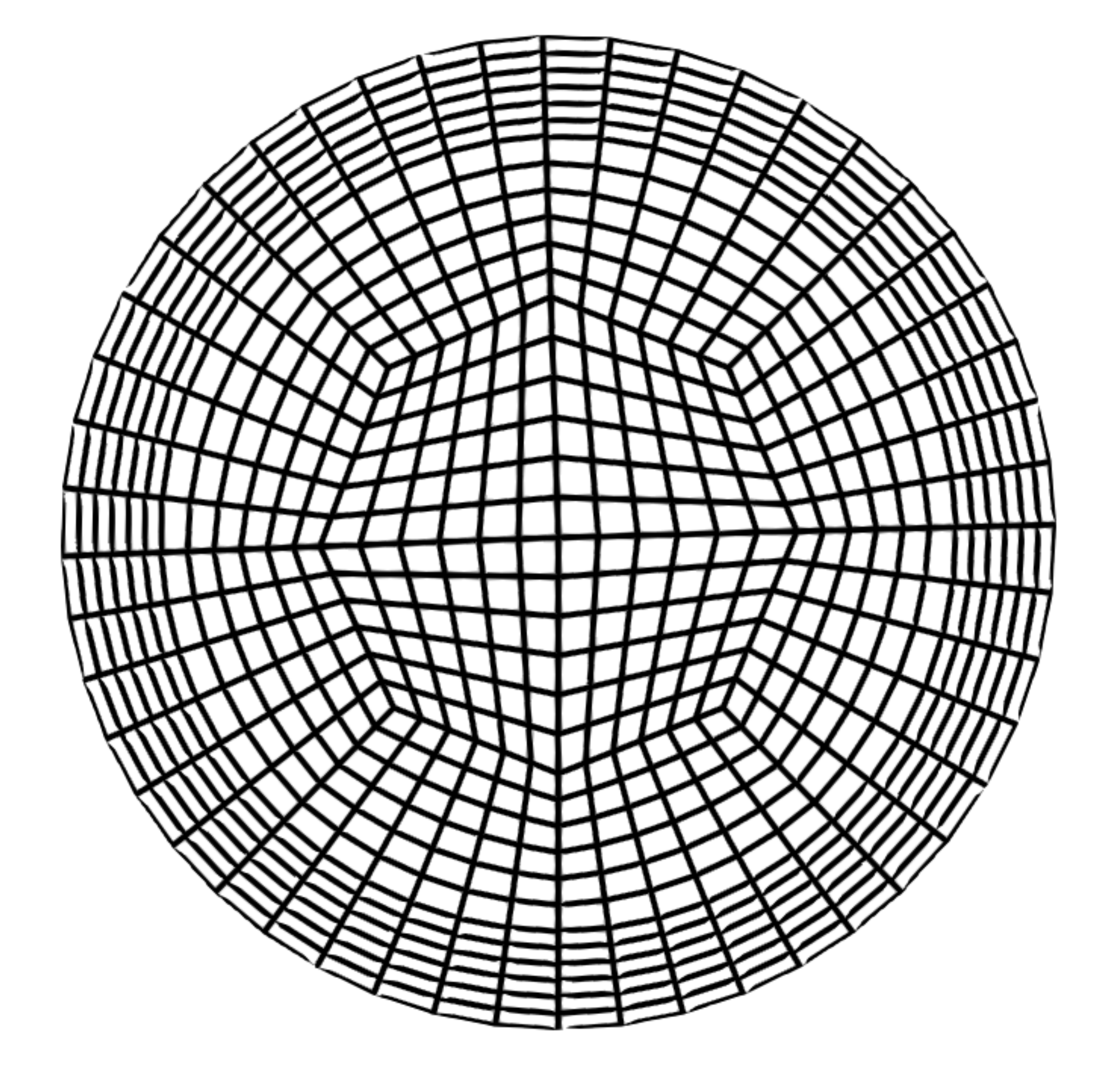}
         \caption{\textit{Left}: Radial cross section depicting radial fidelity at values of 0, 0.5 and 1.}
         \label{fig:left}
     \end{subfigure}\hspace{2em} 
    \begin{subfigure}[b]{0.45\textwidth}
         \centering
         \includegraphics[width=0.32\textwidth]{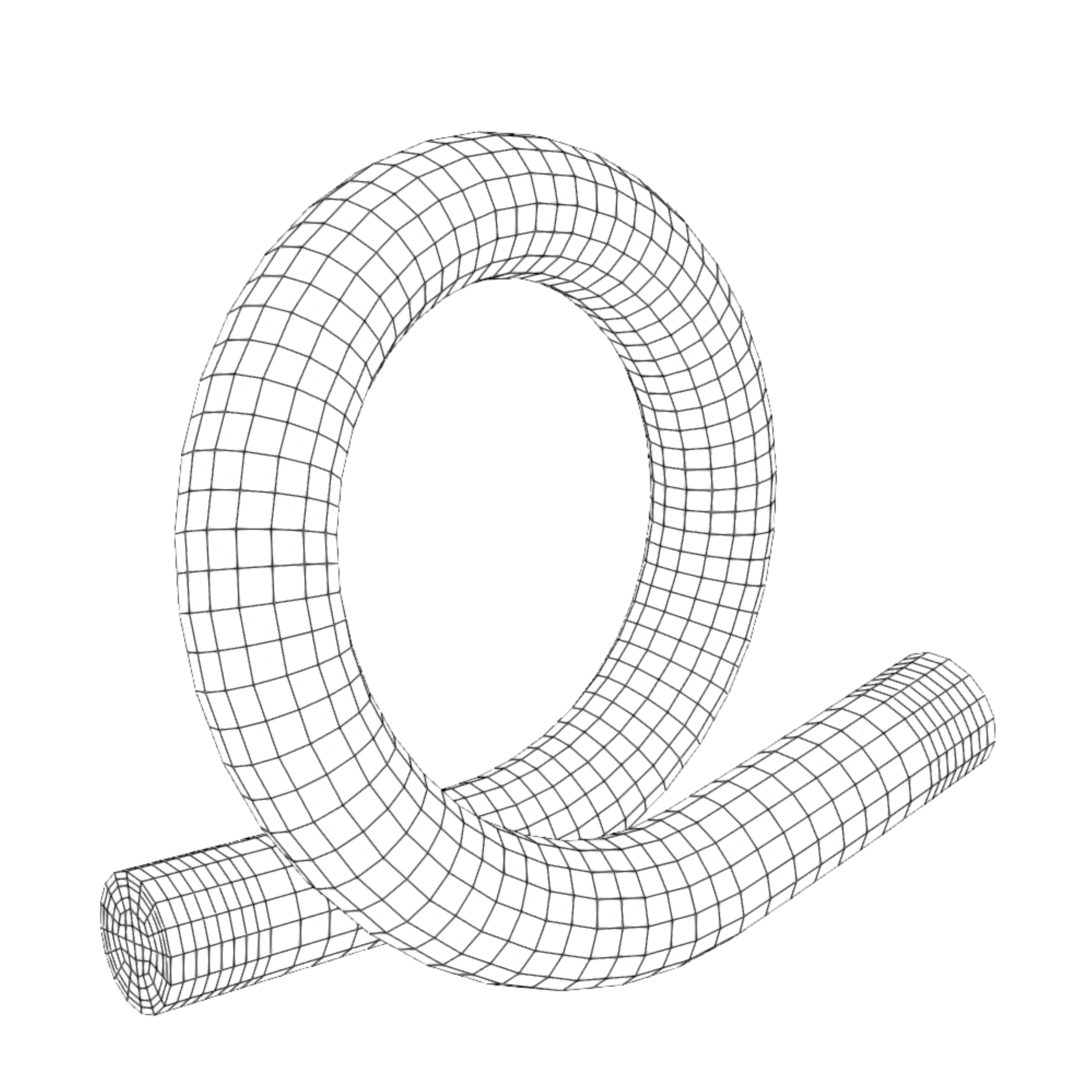}
         \includegraphics[width=0.32\textwidth]{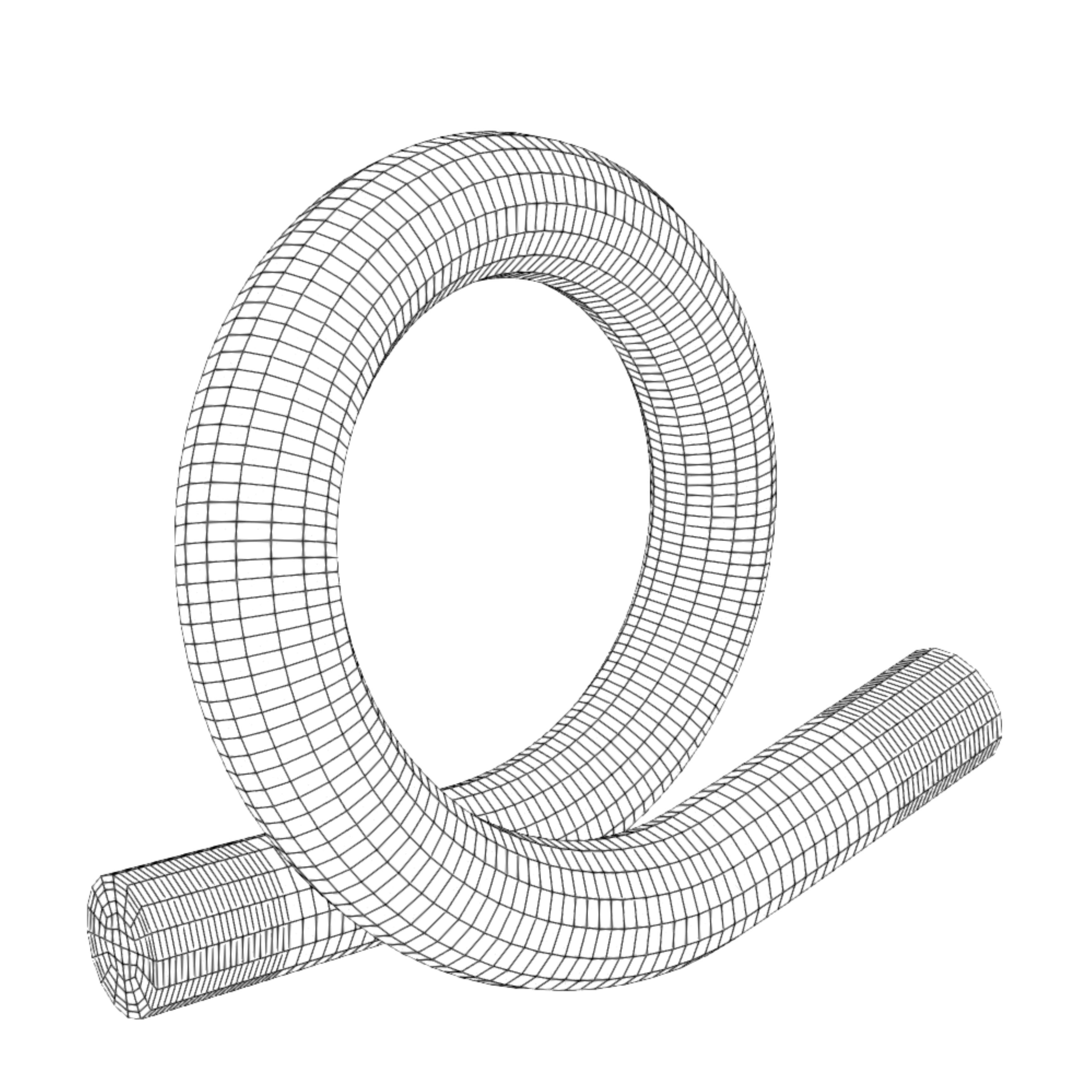}
         \includegraphics[width=0.32\textwidth]{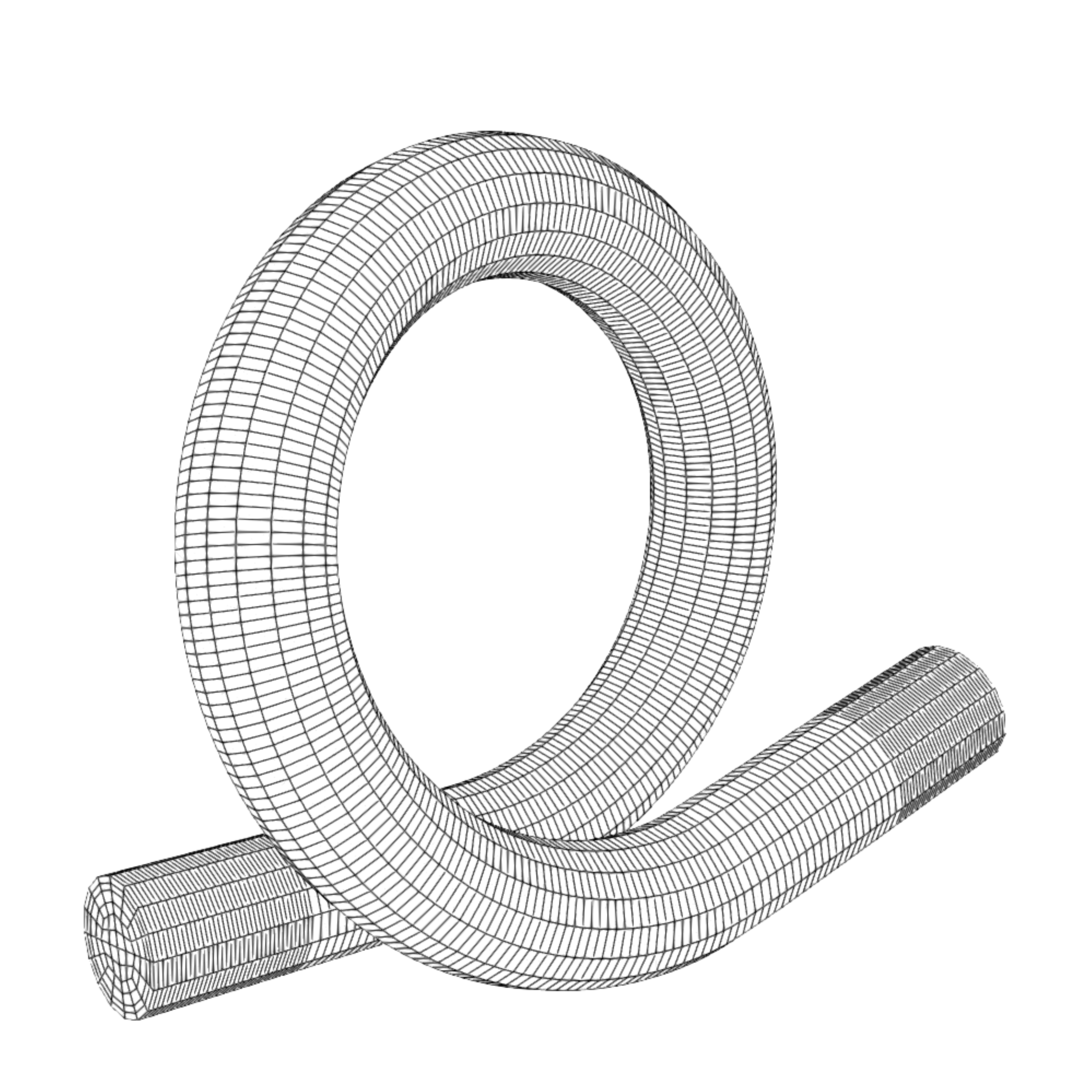}
         \caption{\textit{Right}: View of complete reactor at axial fidelity values of 0, 0.5, and 1.}
         \label{fig:right}
     \end{subfigure}
     \caption{An instance of coiled tube reactor geometry as effected by axial and radial fidelity.}
     \label{fidelities}
\end{figure}

Two fidelity aspects, axial and radial, can be varied when meshing is performed given a tube geometry.
Figure \ref{fidelities} demonstrates how axial and radial fidelity effects the final mesh. 
In this work, we combine both aspects into a single fidelity, and identify five discrete fidelity values.
We leave the case in which axial and radial fidelities are allowed to vary independently for future work.

\begin{figure}[htb!]
    \centering
    \begin{subfigure}[b]{0.45\textwidth}
         \centering
         \includegraphics[width=\textwidth]{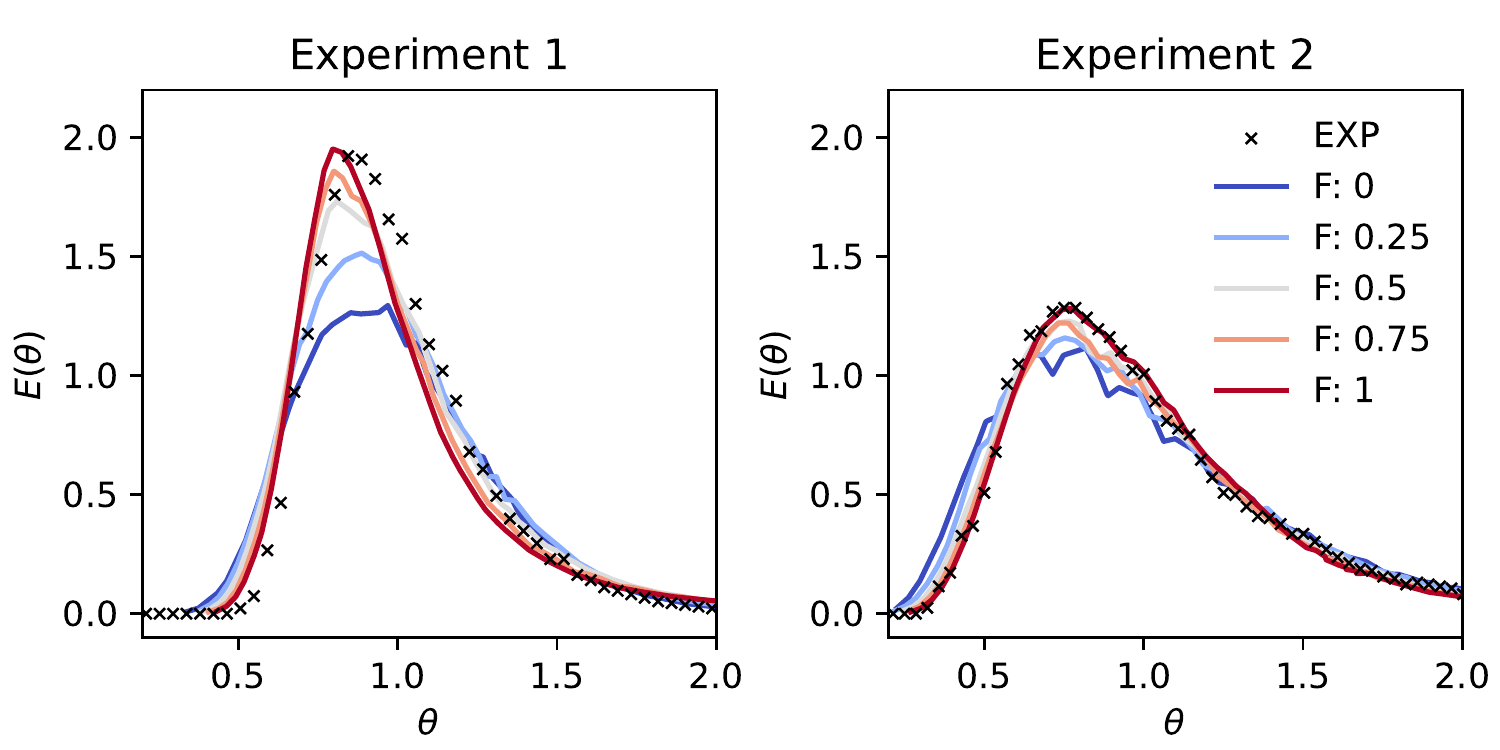}
        \caption{\textit{Left}: The concentration profile of a tracer injection at five fidelity levels between 0 and 1 against experimentally obtained data. $E(\theta)$ represents dimensionless concentration as a function of dimensionless time $\theta$.}   
         \label{fig:left}
     \end{subfigure}\hspace{2em} 
    \begin{subfigure}[b]{0.45\textwidth}
         \centering
         \includegraphics[width=\textwidth]{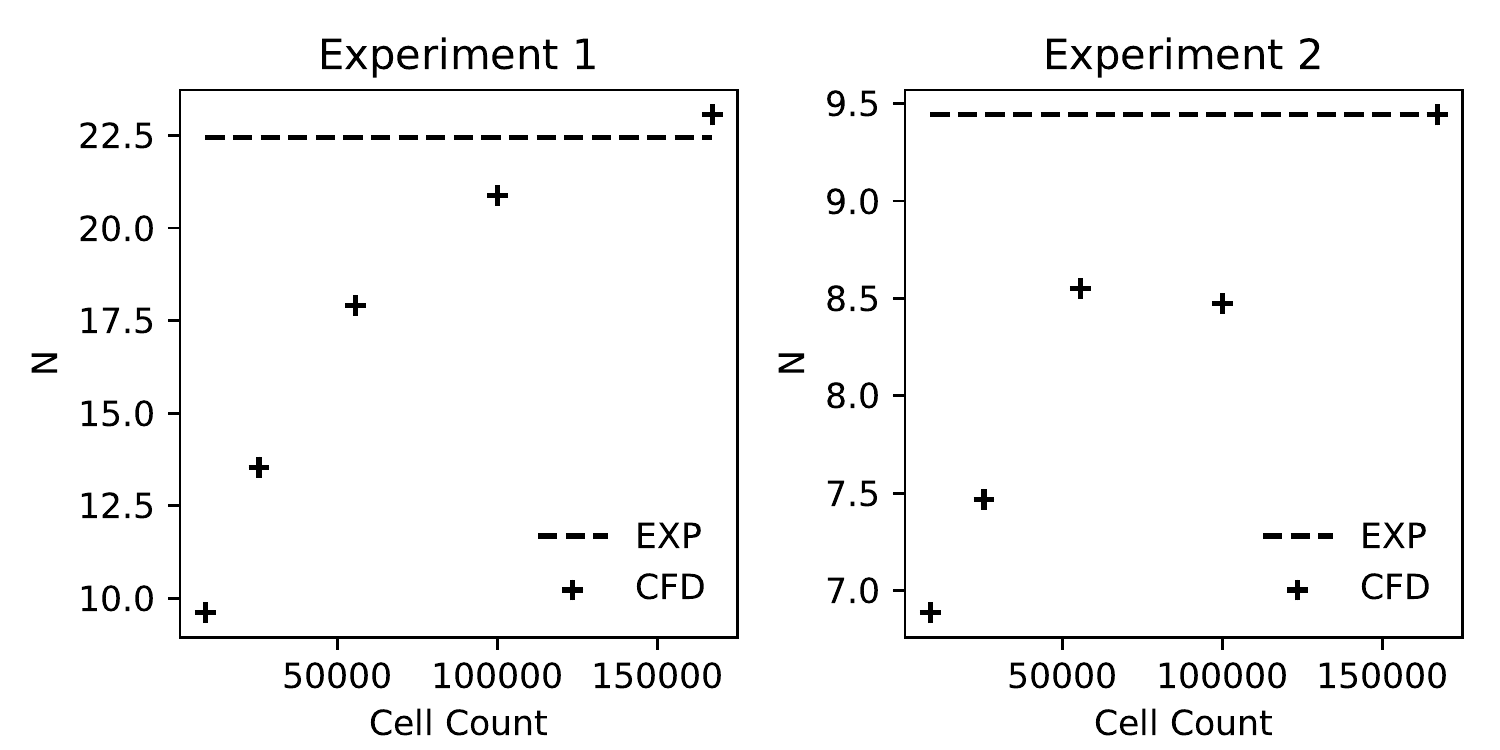}
         \caption{\textit{Right}: The value of N corresponding to the concentration profile from each fidelity, which has been converted to cell count. N represents plug-flow characteristic which is the quantity optimized for.}
         \label{fig:left}
     \end{subfigure}
     \caption{Validation of five discrete mesh fidelities corresponding to different cell counts, across two sets of experimental data under different conditions. }
     \label{validation}
\end{figure}

To investigate the effect of fidelity on function accuracy, the five discrete fidelities were simulated and compared with experimental data.
Figure \ref{validation} validates the tracer concentration profile of simulations using the five fidelities against two sets of experimental data. 
The objective of the optimization is to maximize the plug-flow characteristic $N$ (high values of $N$ are an indicator for good radial mixing and poor axial mixing), which approximately corresponds to fitting the concentration profile with the tank-in-series model \cite{McDonough2019b}.
Figure \ref{validation} also demonstrates how increasing fidelity (and therefore cell count) results in a closer approximation to the experimental value of $N$, derived from each concentration profile.

\subsection{Model specification}
To model each fidelity level we apply Deep Gaussian processes as demonstrated by \citet{https://doi.org/10.48550/arxiv.1903.07320}. 
DGPs provide a natural extension to sequentially trained multi-fidelity models, where a generating function $f$ at a given discrete fidelity $t$ is modeled as a linear or nonlinear function of lower-fidelities plus a mismatch term.
For example in the nonlinear case $f_t(x)$ is given by
\begin{align}
    f_{t}(x) = \rho_{t-1}(f_{t-1}(x),x) + \phi_{t-1}(x).
\end{align}
Multi-fidelity DGPs combine $\rho_{t-1}$ and $\phi_{t-1}$ into a single term $g_{t-1}$ which is modeled using a GP, resulting in a composition of GPs.
For example for $t$ different fidelity levels, and observations $x$ at each fidelity, the highest fidelity is modeled as
\begin{align}
    f_t(x_t,\dots,x_1) = g_t(...g_{t-1}(g_1(x_1),x_{t-1}),x_t).
\end{align}
In this work, we assume five discrete fidelity levels.
Thus the resulting \textsc{DGP} has five layers ($t=5$), corresponding to simulations at fidelities equal to 0, 0.25, 0.5, 0.75, and 1.
MF-DGPs are implemented in Python using \textsc{emukit} \citep{emukit2019}.

\subsection{Multi-fidelity Bayesian Optimization}

We present an approach for multi-fidelity Bayesian optimisation inspired by the MF-GP-UCB algorithm \citep{MFUCB}. 
We find that the number of hyper-parameters in MF-GP-UCB is generally unsustainable for a problem with no prior knowledge, and a large number of fidelities.
Algorithm \ref{MFGPA} demonstrates the simplified approach we apply, inspired by MF-GP-UCB.
The two main differences being the use of DGPs to model relationships between fidelities as opposed to separate models, and a simplified fidelity selection which is directly tied to the computational expense of an evaluation at a given fidelity.
The approach differs from standard UCB-BO by including an additional discrete sub-problem to select the subsequent simulation fidelity, based on uncertainty and computational expense at each fidelity. 

\begin{algorithm}
    \caption{Deep GP-based Multi-fidelity Bayesian Optimization}
    \begin{algorithmic} \label{MFGPA}
    \REQUIRE $f_1(x)\dots f_T(x)$, $\mathcal{X}$, $n$
    \FOR {$t$ in $1,\dots,T$}
        \STATE Generate $n$ samples, $\mathbf{x}_t$, and evaluate $f_t(\mathbf{x})$ resulting in $\mathbf{y}_t$.
        \STATE $\tau_t \leftarrow$ average simulation time 
    \ENDFOR
    \WHILE{\text{Budget not exhausted}}
    \STATE Train DGP using $\mathbf{x}_1,\dots,\mathbf{x}_T$ and $\mathbf{y}_1,\dots,\mathbf{y}_T$ 
    \STATE Solve UCB for highest-fidelity: $x^*\leftarrow \arg\max_x \{\mu_T(x)+\beta^{1/2}\sigma_T(x)|x\in\mathcal{X}\}$
    \STATE Choose fidelity based on variance of DGP and simulation cost: $t^*\leftarrow \text{argmax}_t\{\gamma_t\beta^{1/2}\sigma_t(x^*)\}$ where $\gamma_t = \max(\tau)/\tau_t $
    \STATE Evaluate $f_{t^*}(x^*)$, add $x^*$ to $\mathbf{x}_{t^*}$ and $f_{t^*}(x^*)$ to $\mathbf{y}_{t^*}$
    \ENDWHILE
    \end{algorithmic}
    \end{algorithm}

We note that in choosing $\gamma_t$ based on the points sampled to construct the initial DGP, the algorithm contains no additional hyper-parameters than standard UCB Bayesian optimization.
Additionally $\gamma_t$ may be updated after a simulation has been performed, more accurately reflecting the simulation time at that fidelity. 

\section{Experimental results}

Figure \ref{main_results} demonstrates the MF-DGP based optimization, which shows the selection of multiple simulation fidelities throughout optimization. 

We note that initializing the DGPs for multi-fidelity modeling requires simulations at each fidelity. 
As Bayesian optimization typically demonstrates fast early convergence, this provides a downside over the standard single fidelity approach, as more time is used to generate the initial data set.

\subsection{Recommendations}

Overall we find that it may not be beneficial in situations where a large number of discrete fidelities are available, to apply all fidelities within an optimization framework.
A large number of fidelities results in a large number of hyper-parameters, more difficult to train multi-fidelity models such as DGPs with more layers resulting in longer inference times, and potentially slower exploration. 
We make the recommendation to apply 2 or 3 discrete fidelities in systems with no prior knowledge, despite more being available.
Future work will compare the multi-fidelity approaches with varying number of fidelities with standard approaches and investigate the benefits, as well as apply similar methods on more, industrial case studies for engineering discovery.

\section{Conclusions}

The optimization of coiled tube reactor geometry is critical to maximize plug-flow behavior and investigate industrial viability. The optimization problem is formulated as an expensive black-box problem.
In this paper, we propose multi-fidelity Bayesian optimization using Deep Gaussian processes to find good reactor configurations, taking advantage of less accurate but faster simulations.
We demonstrate experimental validation of five discrete fidelities, and present a modified multi-fidelity Bayesian optimization algorithm which relies on fewer hyper-parameters than existing approaches.
A multi-fidelity DGP provides correct quantification and propagation of uncertainty, which we use to select not only the next experimental design, but also the fidelity of the evaluation within the algorithm. 
Our approach can be extended to other problems involving parameterized CFD simulations.
This work demonstrates an industrially relevant use case of multi-fidelity deep Gaussian processes for the optimization of expensive black-box functions. 
We hope it provides insight and inspiration for people in the machine learning community to develop methods for a variety of applied case-studies and problems.

\section*{Acknowledgements}
The authors would like to acknowledge the funding provided by the Engineering \& Physical Sciences Research Council, United Kingdom through the PREMIERE (EP/T000414/1). Tom Savage would like to acknowledge the support of the Imperial College President's scholarship, and Ilya Orson Sandoval for providing insights for this work. The authors would also like to thank Dr Jonathan McDonough, Newcastle University.

\Urlmuskip=0mu plus 1mu\relax
\bibliographystyle{plainnat} 
\bibliography{main} 

\appendix
\section{Simulation time at each fidelity}
\begin{figure}[htb!]
    \centering
    \includegraphics[width=0.6\textwidth]{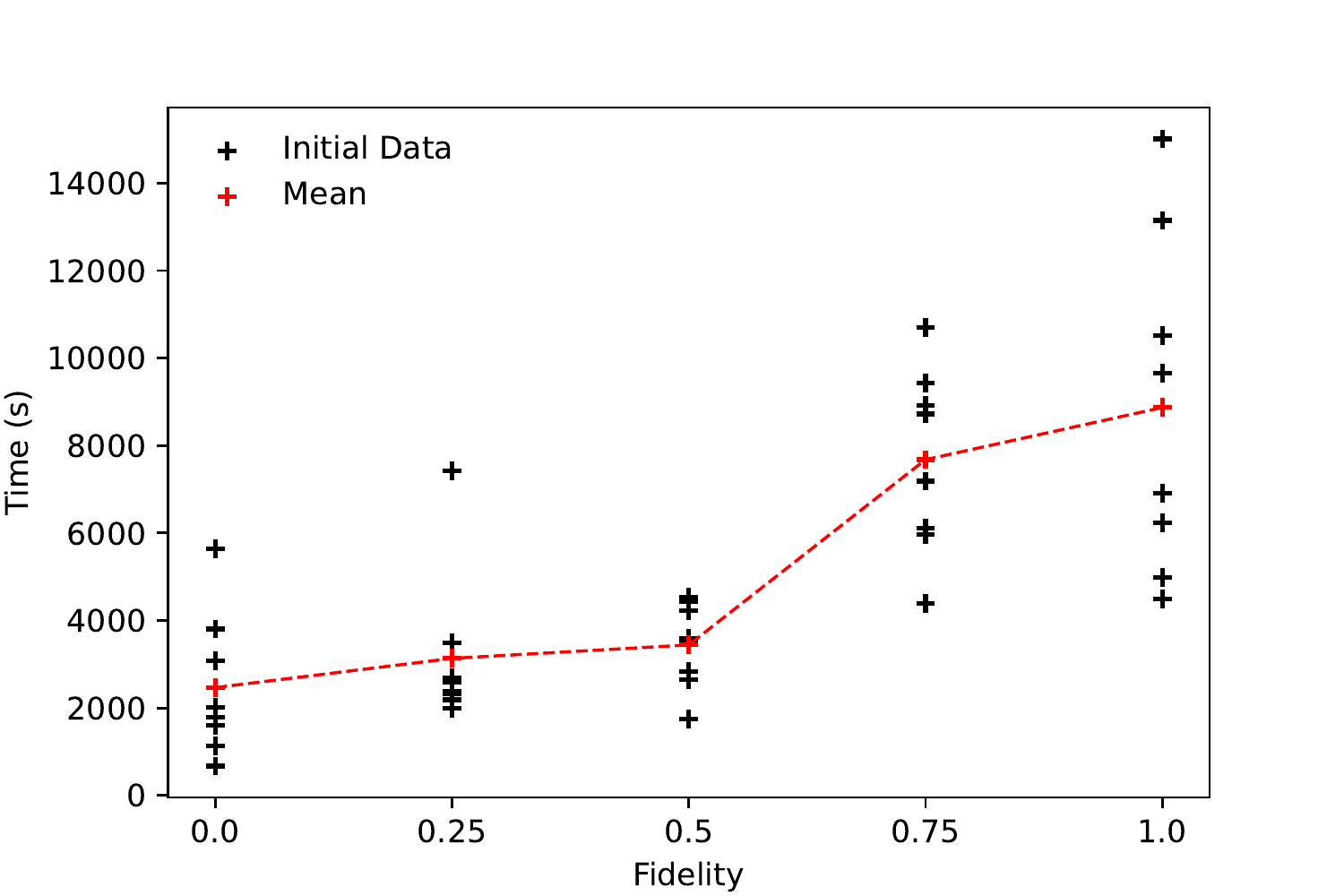}
    \caption{Computational time of simulations at each fidelity across the initial sampled data for multi-fidelity Bayesian optimization.}
    \label{times}
\end{figure}

To dictate hyper-parameters $\gamma$ at each fidelity, weighting the variance of the optimally selected datapoint to select the next fidelity, the mean simulation time is used. 
This value can be updated throughout the optimization, as a more accurate reflection of the computational cost of a simulation is obtained. 
In certain simulations, such as the one we present here, adaptive time-steps are used. 
This results in different simulation times at each fidelity as different reactor geometries will require more or less refined time-steps to simulate. 
It may be possible to model the relationship between parameter space and computational expense in this case, however we leave this for future work. 
Figure \ref{times} shows the individual and mean computational times at each fidelity for the initial datapoints.

\section{Meshing procedure}

\begin{figure}[htb!]
    \centering
    \includegraphics[width=0.6\textwidth]{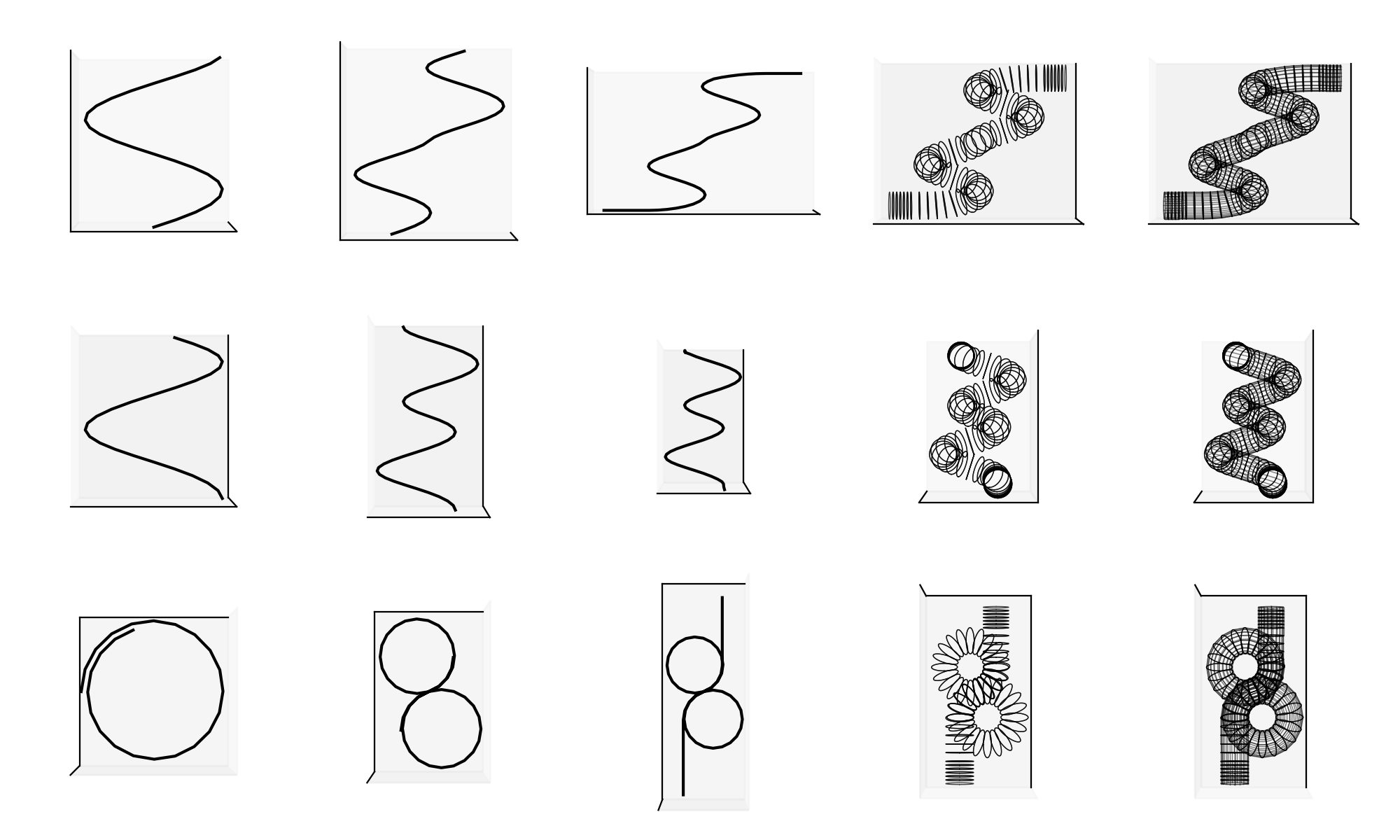}
    \caption{Plots in-line with the X,Y, and Z axis demonstrating the main steps within the plotting procedure for generating a coil given a set of parameters including coil radius, tube radius, pitch, inversion \%, and total volume.}
    \label{meshing}
\end{figure}

Given a set of geometric parameters, meshing was performed in Python using the \textsc{classy\_blocks} library. 
Figure \ref{meshing} provides some insight into how this is performed by plotting the generated mesh structure at each step of the procedure.

\end{document}